\documentclass[12pt, preprint]{aastex}
\usepackage{epsfig}

\begin{document}

\title{The Hanle and Zeeman effects in solar spicules:
a novel diagnostic window on chromospheric magnetism}

\shorttitle{The Hanle and Zeeman effects in solar spicules}

\author{J. Trujillo Bueno\altaffilmark{1,3}, L. Merenda\altaffilmark{1},
R. Centeno\altaffilmark{1}, M. Collados\altaffilmark{1} \&
E. Landi Degl'Innocenti\altaffilmark{2}}
\altaffiltext{1}{Instituto de Astrof\'{\i}sica de Canarias, E-38205, La Laguna,
Tenerife, Spain}
\altaffiltext{2}{Dipartimento di Astronomia e Scienza dello Spazio, Largo Enrico Fermi 2,
I-50125, Firenze, Italy.}
\altaffiltext{3}{Consejo Superior de Investigaciones Cient\'{\i}ficas (Spain)}


\begin{abstract}

 An attractive diagnostic tool for investigating the
 magnetism of the solar chromosphere is the
 observation and theoretical modeling of the
 Hanle and Zeeman effects in spicules,
 as shown in this letter for the first time.
Here we report on spectropolarimetric
observations of solar chromospheric spicules
in the He {\sc i} 10830 \AA\ multiplet and on their
theoretical modeling accounting for
radiative transfer effects. We find that the magnetic
field in the observed (quiet Sun) spicular
material at a height of about 2000 km
above the visible solar surface
has a strength of the order of 10 G and
is inclined by approximately $35^{\circ}$ with respect to the
local vertical direction. Our empirical finding based on
full Stokes-vector spectropolarimetry
should be taken into account in future magnetohydrodynamical
simulations of spicules.

\end{abstract}


\section{Introduction}

 Spicules were described in 1877 by
 Father Angelo Secchi as jet-like, elongated plasma
 structures in the solar
 atmosphere (Secchi 1877).
 These features are best seen when observing a few arcsec
 off the limb in various chromospheric emission lines,
 such as H${\alpha}$ or the lines of
 neutral helium at 5876 \AA\ and 10830 \AA\ .
 It is commonly believed that most of the chromospheric emission
 in these lines comes from spicules, and that at heights
 exceeding 1500 km
 above the photosphere the solar chromosphere is mainly composed
 of spicular material (Beckers 1972). These
 needle-shaped plasma structures
 show apparent upward
 velocities reaching 25 ${\rm km \, s^{-1}}$ lasting for some 5 minutes,
 and are frequently slanted with respect to the solar
 radius vector through the observed point (hereafter,
 solar local vertical).
 After reaching a typical maximum height of 9000 km,
 the ejection stops and is followed by a fading of the spicule brightness
 or a return of the emitting material to the photosphere. Interestingly,
 in the upward-moving phase the spicule mass flux exceeds
 by two orders of magnitude the mass loss of the
 solar corona through the solar wind.

 Practically, all theoretical models aimed at explaining the origin
 of spicules require the presence of
 magnetic fields (e.g., the review by Sterling 2000).
 For instance, in order to model
 dynamic jets in {\em active region} fibrils
 De Pontieu et al. (2004)
 assume a rigid flux tube whose magnetic strength
 changes from 1600 G in the photosphere to 120 G
 in the low corona. What has been really lacking up to now are
 spectropolarimetric investigations to infer the strength and geometry
 of the magnetic field that is thought to channel the spicular motion.
 To fill this gap we started a few years ago an
 investigation which combines spectropolarimetric
 observations and theoretical modeling based on the quantum
 theory of spectral line polarization (see Trujillo Bueno 2003$a$,
 for a brief advance of the results of this investigation).
 In this letter we report on
 a selection of the observed off-limb Stokes profiles
 in the He {\sc i} 10830 \AA\ multiplet,
 which clearly show how
 the Hanle effect (Hanle 1924; Stenflo 1994; Trujillo Bueno 2001)
 rotates the direction of polarization of the scattered light.

\section{Spectropolarimetric observations}

The observations reported here
were carried out on 10 May 2001 with the Tenerife Infrared Polarimeter (TIP;
see Mart\'{\i}nez Pillet et al. 1999)
mounted on the German Vacuum Tower Telescope (VTT) at the Observatorio del Teide (Spain). 
TIP uses ferro-electric liquid
crystal retarders as polarization
modulators. After the light beam is temporally modulated
it goes through a double birefringent plate
that divides it into two orthogonal
polarization beams, which are then imaged on a single detector array.
In order to measure $I, Q, U$ and $V$, TIP takes four
consecutive images with independent analyzer configurations, that result
in linear combinations of the four Stokes parameters.
The information obtained independently from each
polarization beam is combined only at the end of the
data reduction procedure in order to correct for the seeing-induced crosstalk from $I$ to $Q,U$ and $V$.

The spectrograph slit was located at about 2.5 arcsec
off the East solar visible
limb and parallel to it, thus crossing the spicular material
that we could see clearly
in the corresponding H${\alpha}$ slitjaw images.
Note that this off-limb location corresponds approximately
to an atmospheric height of 2000 km
above the visible solar surface, because one has to take into account
that the visible solar limb corresponds to a height of about 250 km
(e.g., Asensio Ramos et al. 2003). It is important to
point out that during our observation
the visible East solar limb region was fairly quiet without any indication of active regions in the slitjaw images. For each fixed slit position
we took various independent time series of 50 consecutive
images, with each of the images resulting from the accumulation
of 5 snapshots of 100 ms.
In order to improve the signal-to-noise ratio we temporally-averaged
the 50 consecutive images of the time series selected,
which implies a net integration time of 209 seconds.

We find that while Stokes $Q$ has a sizable signal that
is always positive at all the spatial points along the
spectrograph slit, the
Stokes $U$ parameter turns out to vary rather smoothly from zero
at the extremes of the spatial domain defined
by the length of the spectrograph's slit ($\sim40$ arcsec long),
to a negative value ($U/I_{\rm max}{\approx}-0.4\%$) at
a spatial point corresponding approximately to
the center of the slit. Our spectropolarimetric observations
of spicules in the He {\sc i} 10830 \AA\ multiplet
are very encouraging,
especially because of the detection of non-zero Stokes $U$ profiles
like the illustrative example shown Fig. 1. According to
the theory of the Hanle effect,
a non-zero Stokes $U$ profile is the
observational signature of the presence of a magnetic field {\em inclined} with respect to the local vertical direction. Finally,
note that the amplitudes of the
Stokes $V$ profiles of the spicules observed on 10 May 2001
were very weak, lying almost at the noise level.

\section{Theoretical modeling of the Hanle and Zeeman
effects in solar spicules}

The determination of the magnetic field vector in solar spicules
can be achieved via theoretical modeling
of the Hanle and Zeeman effects in
suitably chosen spectral lines, such as those
of the He {\sc i} 10830 \AA\ multiplet.
To this end, we have applied the quantum theory
of spectral line polarization,
calculating the
wavelength positions and strengths of the Zeeman components
in the incomplete Paschen-Back effect regime, as
explained in detail in Landi Degl'Innocenti \& Landolfi (2004).
We have assumed that a collection of helium atoms located at
a given height above the visible solar `surface'
is illuminated by the (limb-darkened) photospheric
radiation field, whose center-to-limb variation has been tabulated
by Pierce (2000). The anisotropic radiation
pumping induces population imbalances and quantum
coherences among the magnetic substates of energy levels
(that is, {\em atomic polarization}),
which gives rise to linearly polarized light.
The atomic level polarization
(and the ensuing emergent polarization)
is efficiently modified in the presence of
an inclined magnetic field of strength $B_H{\approx}1.137{\times}10^{-7}/(t_{\rm life}g_L)$, with $B_H$ expressed in gauss and where
$t_{\rm life}$ and $g_L$ are the lifetime (expressed in seconds) and
Land\'e factor of the atomic level under consideration, respectively
(e.g., the review by Trujillo Bueno 2001 on the Hanle effect).


The atomic model we have adopted includes the five lower terms of the
triplet system of helium, namely: $2^3{\rm S}$,  $2^3{\rm P}$,
$3^3{\rm S}$, $3^3{\rm P}$ and $3^3{\rm D}$. The 10830 \AA\
multiplet results from transitions between the metastable
term $2^3{\rm S}$ (which has a single level with total
angular momentum $J=1$) and the term $2^3{\rm P}$
(which has three levels with $J=2,1,0$ in order of increasing energy).
Therefore, it has three spectral lines: a `blue' line at 10829.09 \AA\
(with $J_l=1$ and $J_u=0$) and two `red' lines at 10830.25 \AA\
(with $J_u=1$) and at 10830.34 \AA\ (with $J_u=2$) which appear
blended at the plasma temperatures of solar spicules. The
multiplet that results from transitions between the term
$2^3{\rm P}$ (the upper term of the He {\sc i} 10830 \AA\ multiplet)
and the term
$3^3{\rm D}$ produces the well-known He {\sc i} ${\rm D}_3$
`line' at 5876 \AA\ , which is also of diagnostic
interest (see Section 4)\footnote{Spectropolarimetric observations
of spicules in the ${\rm D}_3$
line have been presented by Sheeley \& Keller (2003),
and also by L\'opez Ariste \& Casini (2005)
in a recently submitted paper.}.

Our interpretation of the spectropolarimetric observations is based
on the solution of the statistical equilibrium equations for the spherical
tensor components (${\rho}^K_Q(J,J')$) of the atomic density matrix
(see the equations of
section {\em 7.6.a} in Landi Degl'Innocenti \& Landolfi 2004).
We have done this by assuming that the helium
atoms (located at ${\sim}2$ arcsec
above the visible solar limb) are radiatively excited by the {\em given}
continuum radiation coming from the underlying solar photosphere,
which is virtually spectrally flat around the wavelengths
of the spectral line transitions that play a significant role
on the ${\rho}^K_Q(J,J')$-values
of the upper and lower terms of the 5876 \AA\ and 10830 \AA\
multiplets\footnote{Under such circumstances, the atomic density
matrix does not depend on the velocity of the helium atoms
in the spicular gas and the complete redistribution theory described
by Landi Degl'Innocenti \& Landolfi (2004) can be safely applied.}.
Such ${\rho}^K_Q(J,J')$ elements allow us to quantify the overall
population of each level of total angular momentum $J$,
as well as the population
imbalances between the magnetic sublevels
pertaining to each $J$-level
and the quantum coherences between pairs of magnetic substates, even between substates pertaining
to different $J$-levels of the same term.
From the calculated density-matrix elements
it is then possible to compute the emission coefficients in the four Stokes parameters,
and the coefficients of the $4{\times}4$
propagation matrix of the Stokes-vector transfer equation
for each of the line transitions
of the assumed multi-term model atom (see
the equations of section {\em 7.6.b} in Landi Degl'Innocenti \& Landolfi 2004).

\subsection{Optically thin modeling}

In a first modeling step we have neglected radiative
transfer effects along the line of sight. This
{\em optically thin} assumption is identical to that
generally adopted for inferring the magnetic field vector
from the Stokes profiles of emission lines observed in solar prominences (see, e.g., Bommier et al. 1994).

The result of our best fit to the observed
Stokes profiles is shown by the solid lines
of Fig. 1. With the exception of Stokes $I$, there is a good
fit to Stokes $Q$, $U$ and $V$ for
a magnetic field vector of strength
$B=10$ gauss, inclination
${\theta}_B=35^{\circ}$ with respect to the
local vertical direction, and
azimuth\footnote{See Fig. 13.1 in Landi Degl'Innocenti \& Landolfi (2004) for the definition
of the angles ${\theta}_B$ and $\chi_B$, and note that
a magnetic field vector lying in the scattering plane has $\chi_B=0^{\circ}$
or $\chi_B={\pm}180^{\circ}$. We have taken
$\delta=0$ in that figure,
which implies that the spicular material is supposed to lie
in the plane of the sky.} $\chi_B=172^{\circ}$.
The inferred magnetic
field vector at those spatial points where the observed
Stokes $U$ was found to be negligible is also inclined by
about $35^{\circ}$, while the azimuth turns out to be
significantly different (e.g., $\chi_B=186^{\circ}$ for a slit point
situated at a distance of 7'' from that of Fig. 1).
The discrepancy found in Stokes-$I$
around the wavelength location of the `blue' component of the
He {\sc i} 10830 \AA\ multiplet (see Fig. 1) indicates that the {\em optically thin}
assumption is not suitable for modeling the Stokes-$I$
profiles of solar chromospheric spicules.

\subsection{Optically thick modeling}

There are various levels of sophistication to account
for radiative transfer effects in solar plasma structures
like prominences, coronal filaments and chromospheric spicules.
Here we consider a relatively simple model with
the basic aim of demonstrating that radiative transfer
effects are indeed at work in solar chromospheric spicules,
but that such effects mainly affect the shape of the
emergent Stokes-$I$ profiles.
To this end, we assume a constant-property slab of
optical thickness ${\tau}$ at the wavelength
under consideration, which accounts for
the collective effect of several individual spicules along
the line of sight. The helium atoms of this slab are
assumed to be polarized as in the previous optically-thin case. 

It is not difficult to show that for this optically-thick case
of a constant-property slab the {\em emergent}
Stokes-$I$ and Stokes-$X$ profiles ($X$ being $Q$, $U$ or $V$)
are given by

\begin{equation}
I({\tau})\,=\,I_0\,{\rm e}^{-{\tau}}\,+\,{{\epsilon_I}\over{\eta_I}}(1\,-\,{\rm e}^{-{\tau}}),
\end{equation}

\begin{equation}
X({\tau})\,=\,X_0\,{\rm e}^{-{\tau}}\,+\,{{\epsilon_X}\over{\eta_I}}(1\,-\,{\rm e}^{-{\tau}})\,-\,
{{{\epsilon_I}{\eta_X}}\over{\eta_I}^2}(1\,-\,{\rm e}^{-{\tau}})\,+\,
{{\eta_X}\over{\eta_I}}{{\tau}}{\rm e}^{-{\tau}}({{\epsilon_I}\over{\eta_I}}\,-\,I_0),
\end{equation}
where $I_0$ and $X_0$ specify the boundary condition
-that is, the Stokes parameters
that illuminate the slab's boundary that is most distant
from the observer. In these expressions
($\epsilon_I$, $\epsilon_X$) are the components of the
emission vector, while ($\eta_I$, $\eta_X$) are
the absorption and dichroism
components of the ($4{\times}4$) propagation matrix.
The approximation we have used to obtain this
analytical solution to the radiative transfer problem in a
constant-property slab is that the general Stokes-vector
transfer equation can be simplified as indicated
by Eqs. (55)--(58) of Trujillo Bueno (2003$b$), which
is indeed justified in our case because in the spicular material
the Zeeman splitting turns out to be a very small fraction of
the spectral line width and also
because at a few thousand kilometers
above the solar
visible `surface' the degree of anisotropy of the
photospheric radiation field is weak (see also S\'anchez Almeida \& Trujillo Bueno 1999).

The boundary condition for modeling
the emergent Stokes parameters from
optically-thick solar spicules observed {\em off-the-limb}
is $I_0=Q_0=U_0=V_0=0$. Therefore,
in contrast with the previously discussed
optically-thin case, we now have that
the slab's optical thickness at the line-core
of the `red line' (${\tau}_{\rm red}$) is the only
{\em additional} free parameter whose value has to be chosen to
fit the observed Stokes profiles.
Figure 2 shows the result of our radiative transfer
modeling of the observed Stokes profiles discussed previously
in Fig. 1. The solid and dotted
lines correspond approximately to the same thermal
velocity ($w_{\rm T}{\approx}14$ km s$^{-1}$), which is now significantly lower than that required to fit the observed spectral line widths via the optically thin modeling. The slab's optical
thickness ${\tau}_{\rm red}$  is also similar in the two modeling cases
corresponding to the solid and dotted lines.
The same happens with
the magnetic field vector which in both cases
turns out to be practically identical to that inferred
via the optically-thin approximation (that is, we now find
$B{\approx}10$ G and $\theta_B{\approx}37^{\circ}$).

The only relevant difference between the two modeling cases
of Fig. 2 is the following.
The dotted lines results from calculations with a
damping constant of the Voigt profile that has not been artificially
enhanced -that is, with
that resulting from the natural
broadening and the assumed `thermal'
velocity, as was the case in Fig. 1.
Interestingly, the corresponding theoretical
Stokes $Q$ profile shows a
tiny negative signal
around the wavelength position of the `blue line' of the
He {\sc i} 10830 \AA\ multiplet. This is nothing but the
observational signature of a differential absorption of
polarization components (dichroism) caused by the presence of a
significant amount of atomic polarization in the ground level
of the triplet system of helium. As seen in Fig. 2,
we obtain a fairly good fit to the observed
Stokes profiles, except in the far wings. We point out that
the fit of the far wings can be improved by
artificially enhancing the damping
parameter of the Voigt profile, as shown by the solid lines, which
might be interpreted as an indication of non-thermal broadening
mechanisms associated with non-maxwellian velocity distribution functions.

Obviously, the presence of
non-thermal broadening mechanisms
makes it difficult to detect the
above-mentioned observational
signature of dichroism
(selective absorption of polarization components),
which results from the presence of
lower-level polarization\footnote{It is also of
interest to mention that when the calculations are carried out
assuming a completely unpolarized ground level, then
the inferred magnetic strength is still $B{\approx}10$ G,
while the inclination of the magnetic field vector is
sligtly smaller (i.e., $\theta_B{\approx}32^{\circ}$).}.
In fact, that negative Stokes $Q$ signal
at the wavelength location of the `blue line'
of the He {\sc i} 10830 \AA\ multiplet
turns out to be a very tiny observational signature
for free-standing slabs with ${\tau}_{\rm red}{<}6$.
As shown by Trujillo Bueno et al. (2002),
the situation is however much more favourable for solar
prominences seen against the bright background
of the solar disk -that is, for the solar filament case
where the boundary condition $I_0{\ne}0$ and
one measures the polarization of the {\em transmitted} beam
after having been selectively absorbed.

It is important to point out that
for magnetic strengths sensibly larger than
10 gauss the He {\sc i} 10830 \AA\ multiplet enters into the saturation
regime of the upper-level Hanle effect where the Stokes $Q$ and $U$
parameters are only sensitive to the orientation of the magnetic field
vector.  For this reason, it is crucial to measure also Stokes $V$,
as we have done in this investigation,
since the observed amplitude allows us to estimate in a rather
straightforward way
how large the magnetic strength can be.
Indeed, for magnetic strengths weaker than the
crossing field of the $J$-levels of the upper term (${\sim}400$ G)
the circular polarization of the He {\sc i} 10830 \AA\ multiplet
is dominated by the Zeeman splitting, instead of by the {\em alignment-to-orientation} mechanism discussed by Landi Degl'Innocenti \& Landolfi (2004). 
We have carried out several model calculations
of the emergent spectral line polarization
for increasing values of the magnetic field strength,
paying particular attention to compare the calculated and observed
circular polarization amplitudes.
As a result, we have found that the best fit
to the observed (temporally-averaged) Stokes profiles
is obtained for $B=10$ G, and that magnetic strengths
sensibly larger than 15 gauss
would be incompatible with the (quiet Sun) chromospheric spicules
we observed on 10 May 2001. It is however important to note
that the observed Stokes profiles also include
the unavoidable averaging along the line of sight. Obviously, we cannot
exclude the possibility of stronger fields occupying only
a small fraction of the integration volume along the line of sight.

Finally, it is of interest
to mention that the measured circular polarization was also very weak
for the off-limb spicules we observed during September 2003. However,
some of the chromospheric spicules we have observed during
September 2004 showed sizable Stokes $V$ signals which
seem to be compatible with an inclined magnetic field of strength
$B{\approx}38$ G. This suggests that the
magnetic field strength of solar spicules can also be significantly
larger than 10 G (see also L\'opez Ariste \& Casini 2005).

\section{Conclusions}

The reported spectropolarimetric
observations of solar chromospheric spicules
in the He {\sc i} 10830 \AA\ multiplet show clearly
the observational signature of the Hanle
effect (see the non-zero Stokes $U$ profile of Fig. 1),
which provides the first {\em direct} empirical
demonstration that the spicular
material is significantly magnetized.

In order to obtain information on the strength and geometry
of the magnetic field vector we have applied
the quantum theory of spectral line polarization at two levels of sophistication:
optically thin and optically thick modeling. This has allowed us
to demonstrate that radiative transfer effects have to be taken
into account for a correct modeling of the observed
Stokes-$I$ profiles, and that such transfer effects reduce the
value of the thermal velocity needed to fit the
spectral line widths. Our spectropolarimetric observations
of (quiet Sun) chromospheric spicules
indicate the presence of significantly inclined magnetic fields,
with inclination angles similar to those of the observed spike-like
features themselves (i.e., $\theta_B{\approx}35^{\circ}$). The
magnetic field strength of the (quiet Sun) spicular material
we observed on 10 May 2001 is about 10 G. We think that 10 G is
the typical value for the magnetic strength of the (quiet Sun)
spicular material at an atmospheric height of 2000 km, but
significantly stronger fields may also be present.

An interesting investigation for the near future concerns
the height variation of the magnetic field
that channels the spicular motions,
with particular interest on determining whether or not
it is twisted around the axis of the spicules. To this end,
co-spatial and simultaneous spectropolarimetry
in the 10830 \AA\ and D$_3$ multiplets of neutral helium
would be the most suitable ground-based diagnostic
window. On the one hand, while the linear polarization of the 10830 \AA\ multiplet is sensitive (via the {\em upper-level} Hanle effect) to magnetic strengths between 0.1 and 10 gauss,
approximately, the sensitivity range
for the ${\rm D}_3$ line lies between 0.7 and 70 gauss.
On the other hand, such simultaneous observations
would allow us to avoid a subtle ambiguity,
which is different from the well-known $180^{\circ}$ ambiguity mentioned
in the figure legends. In reality, for magnetic field inclinations
$\theta_B$ such that 
${\theta}_1<{\theta}_B<{\theta}_2$ (with $\theta_1{\approx}30^{\circ}$
and $\theta_2{\approx}150^{\circ}$
when we are in the  Hanle-effect saturation regime),
the magnetic field vector inferred
from the observed polarization in the He {\sc i} 10830 \AA\ multiplet
has an additional ambiguity for some values of the magnetic field
azimuth (see Merenda et al. 2005).
This $90^{\circ}$ ambiguity in the plane of the sky, also known
as the Van Vleck ambiguity,
was pointed out by House (1977) and Casini \& Judge (1999)
concerning the scattering polarization in forbidden coronal lines.
We should mention that from the two possible magnetic field orientations that produce similar Stokes profiles,
in this paper we have always chosen that which lies closest
to the observed inclinations of spicules,
because of the argument that the
observed spicular motions are likely channelled by the magnetic
field vector.

\acknowledgments
We thank Roberto Casini (HAO)
and Arturo L\'opez Ariste (THEMIS)
for scientific discussions.
This research has been funded by the
Spanish Ministerio de Educaci\'on y Ciencia
through project AYA2004-05792 and by the European
Solar Magnetism Network.


\begin{figure}
\plotone{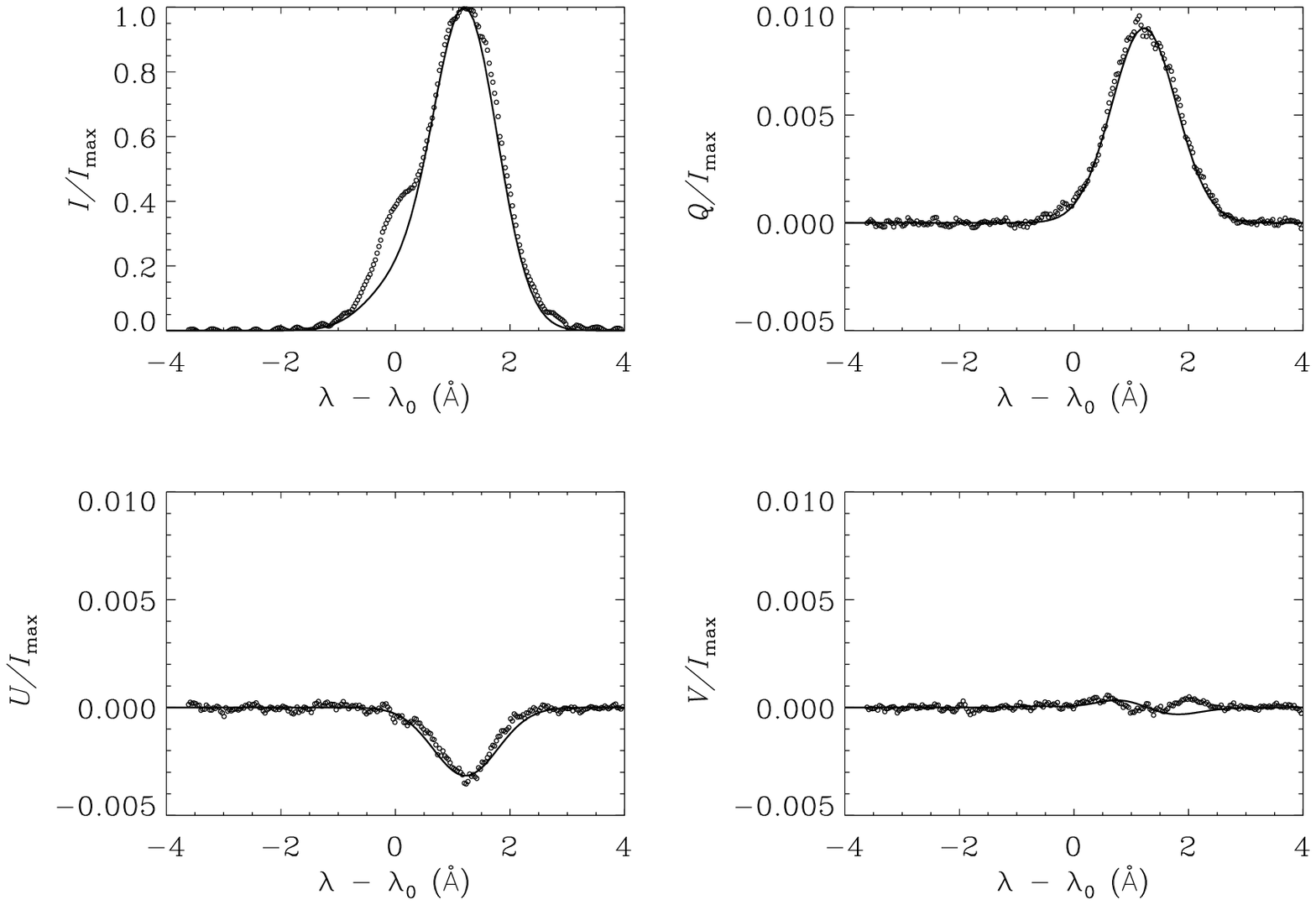}
\caption{Open circles: the observed Stokes profiles
at one of the spatial points that show non-zero Stokes-$U$
signals in the observed
(quiet-Sun) chromospheric spicules.
The reference direction for Stokes $Q$ is
the parallel to the solar limb. The origin of the
wavelength scale corresponds to the blue component of the
He {\sc i} 10830 \AA\ multiplet. Solid line: optically thin theoretical
modeling for strength $B = 10$ G, inclination
$\theta_B = 35^{\circ}$, azimuth $\chi_B = 172^{\circ}$
and a thermal velocity of 22 km${\rm s}^{-1}$.
The alternative
determination $B = 10$ G, ${\theta_B}^{'} = 180^{\circ}-{\theta_B}$, ${\chi_B}^{'} = -{\chi_B}$, gives the same
theoretical Stokes profiles. See footnote 6 for the definition
of the angles $\theta_B$ and ${\chi_B}$}.
\end{figure}

\begin{figure}
\plotone{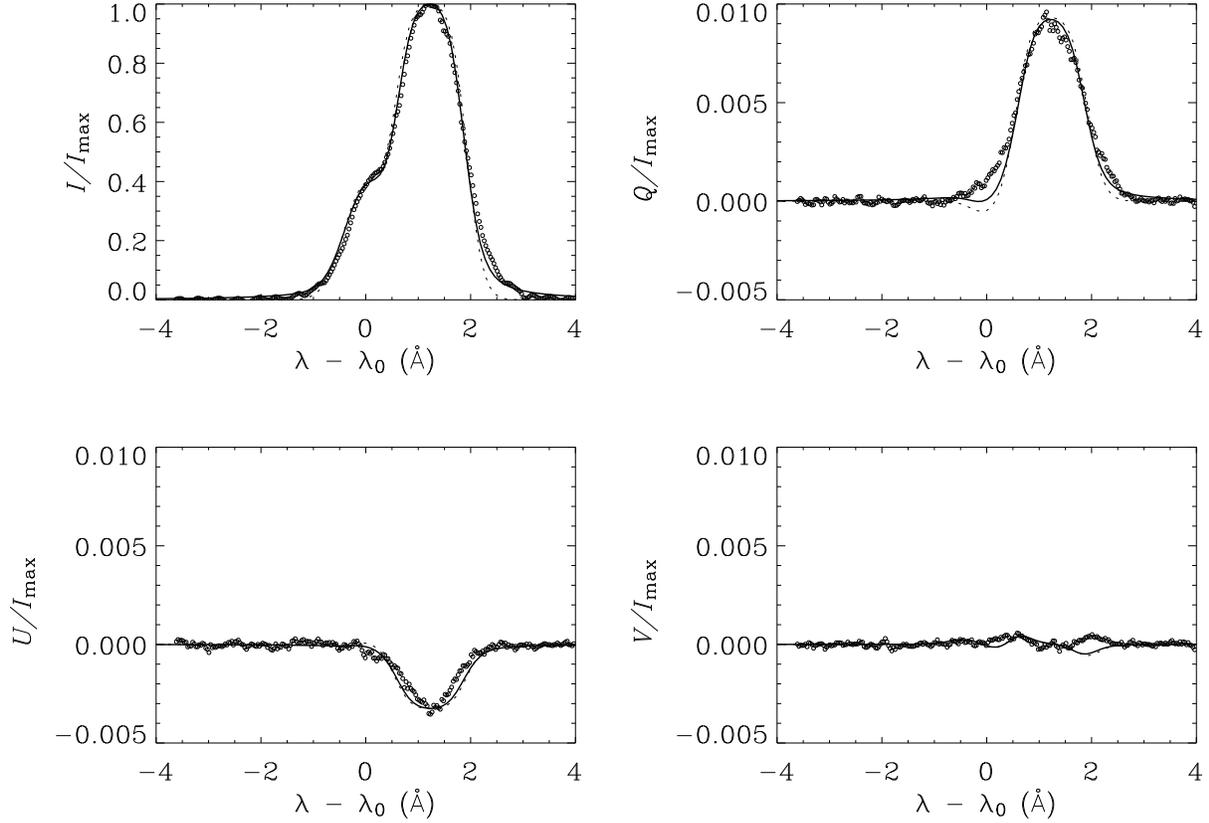}
\caption{Open circles: the observed Stokes profiles
at the same spatial point of Fig. 1. 
The reference direction for Stokes $Q$ is
the parallel to the solar limb.
The origin of the
wavelength scale corresponds to the blue component of the
He {\sc i} 10830 \AA\ multiplet.
Dotted line: optically thick theoretical
modeling (${\tau}_{\rm red} = 3.7$) for a magnetic field strength $B = 10$ G, inclination
$\theta_B = 37^{\circ}$, azimuth $\chi_B = 173^{\circ}$
and a thermal velocity of 15 km${\rm s}^{-1}$. Solid-line:
optically thick theoretical
modeling (${\tau}_{\rm red} = 3$) with enhanced damping
parameter,
for a magnetic field strength $B = 10$ G, inclination
$\theta_B = 37^{\circ}$, azimuth $\chi_B = 173^{\circ}$
and a thermal velocity of 13.5 km${\rm s}^{-1}$.
In both modeling cases, the alternative
determination $B = 10$ G, ${\theta_B}^{'} = 180^{\circ}-{\theta_B}$, ${\chi_B}^{'} = -{\chi_B}$, gives the same
theoretical Stokes profiles.}
\end{figure}

\end{document}